# Differentially Private Location Privacy in Practice

Vincent Primault*, Sonia Ben Mokhtar*, Cédric Lauradoux† and Lionel Brunie*
*Université de Lyon, CNRS
INSA-Lyon, LIRIS, UMR5205, F-69621, France
{vincent.primault,sonia.ben-mokhtar,lionel.brunie}@liris.cnrs.fr
†INRIA
cedric.lauradoux@inria.fr

*Abstract*—With the wide adoption of handheld devices (e.g., smartphones, tablets), a large number of location-based services (also called LBSs) have flourished providing mobile users with real-time and contextual information on the move. Accounting for the amount of location information they are given by users, these services are able to track users wherever they go and to learn sensitive information about them (e.g., their points of interest including home, work, religious or political places regularly visited). A number of solutions have been proposed in the past few years to protect users location information while still allowing them to enjoy geo-located services. Among the most robust solutions are those that apply the popular notion of *differential privacy* to location privacy (e.g., *Geo-Indistinguishability*), promising strong theoretical privacy guarantees with a bounded accuracy loss. While these theoretical guarantees are attracting, it might be difficult for end users or practitioners to assess their effectiveness in the wild. In this paper, we carry on a practical study using real mobility traces coming from two different datasets, to assess the ability of Geo-Indistinguishability to protect users' points of interest (POIs). We show that a curious LBS collecting obfuscated location information sent by mobile users is still able to infer most of the users POIs with a reasonable both *geographic* and *semantic* precision. This precision depends on the degree of obfuscation applied by Geo-Indistinguishability. Nevertheless, the latter also has an impact on the overhead incurred on mobile devices resulting in a privacy versus overhead trade-off. Finally, we show in our study that POIs constitute a quasi-identifier for mobile users and that obfuscating them using Geo-Indistinguishability is not sufficient as an attacker is able to re-identify at least 63 % of them despite a high degree of obfuscation.

## I. INTRODUCTION

More and more people carry handheld devices every day (e.g., smartphones, tablets) equipped with localisation capabilities (e.g., embedded GPS chips, localisation using Wi-Fi/3G/4G hotspots/cells), allowing them to access a wide variety of online services on the move. These services, often called location-based services (or LBSs for short) provide users with a variety of functionalities such as the access to contextual information (weather forecast, road traffic), the ability to find friends or points of interests in the neighbourhood (e.g., Foursquare, Google Maps) and even to play social games (e.g., Map of the Dead[1]). However, to obtain the desired service, the user must communicate to the LBS her current location. As shown in [1], [2], this raises severe privacy concerns due to the knowledge LBSs are able to infer about users. For instance, data mining techniques may allow a curious LBS to find home and work places of an individual, model its mobility patterns with Markov chains, predict its next movement, learn about its political or religious preferences or to infer the users social network by simply using geolocated mobility traces at its disposal. If this information falls between the hands of malicious adversaries, the users can even be exposed to serious threats (e.g., robbers can obtain a list of users that are currently away from home[2]).

Consequently, a number of solutions have been investigated in the literature to protect a user's location privacy while still allowing her to query LBSs for information [3]–[6]. However, most of these solutions either suffer from inaccurate results (e.g., by obfuscating real locations [3]), generate a large amount of traffic in the network (e.g., by generating dummy requests [4]) or poorly integrate with the LBS (see [5]). Recent work proposed in [6], relies on the *differential privacy* principle [7] adapted to the protection of location data. In this solution, called Geo-Indistinguishability (Geo-I for short), location information is protected by adding a controlled amount of noise according to the desired level of privacy, while allowing to get useful information from a LBS with a bounded loss of results. Another key feature of Geo-I is that it does not require changes on the LBS as noise is added on the mobile device. This makes Geo-I a promising solution towards the protection of users location information.

While differential privacy in general and Geo-I in particular provide attracting theoretical guarantees, it might be difficult for practitioners or end users to get a feeling of the practical guarantees it offers. The objective of this paper is to study the effectiveness of Geo-I, in practice. Specifically, we study the ability of Geo-I to protect users points of interests (POIs). Towards this purpose, we use two real mobility traces, i.e., a trace of 536 cabs collected in the city of San Francisco [8] and a trace of 60 users collected in the city of Beijing [9]. For each user in these traces we start by identifying their POIs using a spatio-temporal clustering algorithm inspired from [2], [10], [11]. We refer to these POIs as the *real POIs*. We then obfuscate the mobility traces w.r.t. Geo-I and consider this information as the one collected by the adversary. We assume the latter applies the same clustering algorithm on the obfuscated trace in order to infer the *obfuscated POIs*. Follows our study of the amount and the accuracy of information obtained by the adversary using a set of metrics. These metrics include: the number of real POIs the adversary was able to infer (i.e., *recall*), the physical distance between real POIs and obfuscated ones (e.g., *geographic distance*) and the similarity between neighbourhood surrounding real POIs and obfuscated ones (i.e., *semantic distance*). Our results demonstrate that whatever the level of obfuscation applied by the Geo-I, a curious LBS is able to associate at least 60 % of the obfuscated POIs with real ones in both datasets. Nevertheless, the accuracy of these POIs in terms of geographic and semantic distance depends on the level of obfuscation applied by Geo-I, i.e., the higher the noise the lower the precision of the inferred information. Yet, by studying the overhead generated by Geo-I, we demonstrate that with a high level of obfuscation, mobile users have to filter up to 90 % of inaccurate replies sent from the LBS. Finally, we carried out a re-identification attack where a curious LBS that gathered a set of anonymised POIs using Geo-I for each user, tries to re-identify known users characterised by their set of real POIs. Results show that such curious LBS is able to re-identify from 63 % to 89 % of users according to the degree of obfuscation applied by Geo-I. This result shows that POIs constitute a quasi-identifier and obfuscating them using Geo-I is not sufficient to protect users from re-identification.

---
[1] http://www.mapofthedead.com

[2] http://pleaserobme.com/

The remaining of the paper is structured as follows: we start by presenting an overview of our study in Section II. We then describe the algorithm used for extracting the POIs in Section III and the obfuscation algorithm in Section IV. Further, we present our evaluation metrics in Section V and the results of our study in Section VI. We finally expose some related work in Section VII before to conclude the paper and present our future research directions in Section VIII.

## II. OVERVIEW OF THE STUDY

The goal of this paper is to study the behaviour of differentially private mechanisms for the protection of location information with a specific focus on the Geo-I system as a representative protection. Note that we do not aim at analysing the theoretical guarantees of Geo-I as the latter have already been formally proven by the authors in [6]. Instead, we aim at practically evaluating the degree of protection offered by Geo-I if used by users to obfuscate the location information they send to a curious LBS.

To achieve this objective, we consider users equipped with GPS-enabled devices and who query LBSs on the move to locate remarkable places around them, i.e., *features*, such as point of views, restaurants or subway stations. We suppose LBSs are honest-but-curious, i.e. they answer accurately but want to collect knowledge about users by analysing places they visit. As demonstrated in [12], most of mobile applications have the permission to locate a user and to access the network at any time. This allows curious LBSs to collect and send users' positions periodically and not only when the user queries the LBS. This position is timestamped and linked with a unique user identifier, allowing the LBS to build a *mobility trace* of each single user. We focus in this paper on the ability of Geo-I to protect users points of interest (POIs for short). Indeed, it has been shown that the inference of POIs constitutes a severe privacy breach [1], [2]. For instance, inferring a user POI may allow a curious LBS to find a user's home and work place or even to infer her religious and political preferences if she regularly visits a religious site or a party head-quarter. It can ultimately lead to user de-anonymisation by using reverse geocoding to put a name on visited places and using uniqueness of mobility patterns. It is also the first step of other privacy attacks like next movement prediction [13].

In this paper, we define a POI as the centroid of an area where a user frequently spends a given amount of time. The size of the area, expected frequency and duration of stay in a given place that are required to characterise a POI are parameters of our study. Considering this definition of POIs, our study decomposes into the following four steps:

1) the extraction of POIs from the real mobility traces. We consider these POIs as the ground truth of our study and call these POIs the *real POIs*. We describe our POIs' extraction algorithm in Section III;
2) the obfuscation of the mobility trace using Geo-I. This step is described in Section IV;
3) the extraction of POIs from the obfuscated trace using the same algorithm used to extract the real POIs. We call these POIs the *obfuscated POIs*; and finally
4) an analysis of the accuracy of the obfuscated POIs compared to the real ones. At the heart of this analysis resides the definition of a set of metrics enabling to quantify the level of privacy users get and the loss of precision incurred by running the corresponding obfuscation algorithm. Our metrics are defined in Section V while the results of the analysis are described in Section VI.

An example illustrating the objectives of our study is depicted in Figure 1. In this Figure, a mobility trace as it may be collected by a LBS tracking a user moving in Paris is represented by a set of dots (each dot represents an individual timestamped location). Note that timestamps are not shown for the sake of readability. The POIs that can be extracted from this trace are represented in the same figure as the centroids of the circles surrounding groups of dots. If the user applies Geo-I to obfuscate her individual locations when sending her queries to the LBS, the latter gets the obfuscated trace depicted on the same Figure and represented by a set of diamond shapes. As a result, the LBS would infer a set of obfuscated POIs represented by the dotted circles surrounding groups of diamond shapes. Our objective throughout the paper is to study the accuracy of the obfuscated POIs compared to the real ones.

## III. EXTRACTING POINTS OF INTEREST

To extract users' POIs from a mobility trace, our algorithm, which is depicted in Algorithm 1, decomposes in two main parts. The first part (lines 1-20) allows the extraction of places in which the user spent a certain amount of time, i.e., *stays* as defined in [11]. Identifying stays requires to fix the following two parameters:

1) a time threshold $minTime$, which represents the minimum time that has to be spent in every stay;
2) a distance threshold $maxDistance$ representing the maximal diameter of the stay area.

The threshold $minTime$ depends on the purpose of the extraction algorithm. Indeed, one can consider short stays (e.g., to identify visits to shopping malls) or long stays (e.g., to identify holiday periods). In its first part, our algorithm groups points composing a mobility trace (stored in list $points$ ordered by time) into a set of stay areas. To do so, it builds successive candidate stays; the current one is stored as an ordered list of points in $candidate$. Our algorithm iterates over each point (line 4) and tests if by adding this new point to candidate stay, the diameter of the latter remains under the $maxDistance$ threshold (line 5-6). By convention, if $candidate$ is empty, the latter test succeeds. If not satisfied (line 9), we check if elapsed time inside candidate stay is above the $minTime$ threshold. If so, candidate stay is valid and added to $stays$ (line 11); a new candidate stay is created (line 12). If not, we are not able to create a valid stay. Therefore, we remove the first element of our candidate stay (line 14) and try again to add current point at the next step.

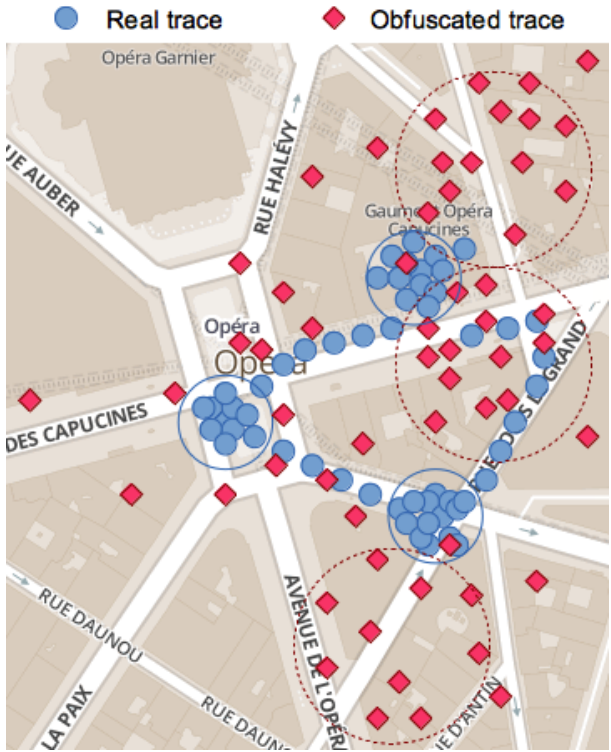

Figure 1. Example of a mobility trace and its obfuscated version.

In order to merge frequent and nearby stay areas, we use in the second part of our algorithm the DJ-clustering algorithm introduced in [10] (lines 21-33). This algorithm creates clusters with a minimal number of points and at a maximal distance to other clusters. In its original version, this algorithm uses a preprocessing step to filter out static points (i.e., points where the speed of the user is zero). In our case we do not consider this step because we are already working on stays.

We run this algorithm on the centroids of stay areas instead of the areas themselves to have something easier to manipulate. This algorithm relies on two parameters:

1) a merge threshold which defines the distance before two distinct clusters are merged in a single one. It is defined as a function of $maxDistance$ threshold in our algorithm;
2) a minimum number of points $minPts$ necessary to form a cluster. This is the number of different stays we have inside a cluster (modulo the distance threshold) to maintain the latter. It gives us the notion of frequency of apparition of a stay and helps to eliminate "accidental" stays that occur only once.

We set the merge distance threshold as a function of the stay distance threshold used in the first step, i.e., 75 % of the latter in order to merge nearby clusters while limiting the risk of creating artificial clusters not representing any ground truth.

As said before, POIs extracted from the original mobility trace using the algorithm we have just presented are considered as ground truth.

---

**Algorithm 1** Algorithm used to extract POIs.

**Require:** $minTime > 0, maxDistance > 0, minPts \geq 1, |points| > 0$
1: $stays \leftarrow \varnothing$
2: $candidate \leftarrow \varnothing$ {ordered list of points}
3: $i \leftarrow 0$
4: **while** $i < |points|$ **do**
5:    $diameter \leftarrow max(dist(points[i], p) \; \forall p \in candidate)$
6:    **if** $diameter \leq maxDistance$ **then**
7:       add $points[i]$ to $candidate$
8:       $i \leftarrow i + 1$
9:    **else**
10:      **if** elapsed time in $candidate \geq minTime$ **then**
11:         add centroid of $candidate$ to $stays$
12:         $candidate \leftarrow \varnothing$
13:      **else**
14:         remove first element of $candidate$
15:      **end if**
16:    **end if**
17: **end while**
18: **if** elapsed time in $candidate \geq minTime$ **then**
19:    add centroid of $candidate$ to $stays$
20: **end if**
21: $clusters = \varnothing$
22: **for** $stay$ in $stays$ **do**
23:    $neighborhood \leftarrow \{s \in stays \text{ s.t. } dist(s, stay) \leq maxDistance \times 0.75\}$
24:    **if** $|neighborhood| \geq minPts$ **then**
25:      **for** $cluster$ in $clusters$ **do**
26:         **if** $neighborhood \cap cluster \neq \varnothing$ **then**
27:            $neighborhood \leftarrow neighborhood \cup cluster$
28:            remove $cluster$ from $clusters$
29:         **end if**
30:      **end for**
31:      add $neighborhood$ to $clusters$
32:    **end if**
33: **end for**
34: **return** centroids of $clusters$

---

## IV. LOCATION-PRIVACY PROTECTION MECHANISM

Literature contains many solutions aiming to protect location information of mobile users. We focus on propositions based on location perturbation, i.e., adding calibrated noise to each location before sending it to a LBS in order to hide the user's real location. These mechanisms seem to be the most promising because they can be applied locally without the need of a trusted third-party and can be integrated with any existing LBS. We focus in this paper more specifically on approaches that rely on *differential privacy*, although a similar study can be conducted on other noise-based protection mechanisms.

## A. Differential privacy

Differential privacy is a concept introduced by Dwork in [7]. It defines formal privacy guarantees applied to statistical databases. The idea is that an aggregate result over a database should be the same whether or not a single element is present inside the database or not. In other words, the addition or removal of one single element does not change significantly the probability of any outcome of aggregate functions.

*Definition 1:* A randomized mechanism $\mathcal{K}$ gives $\epsilon$-differential privacy if for every databases $D_1$ and $D_2$ differing on at most one element and for every $S \subseteq Range(\mathcal{K})$, we have:

$$Pr[\mathcal{K}(D_1) \in S] \leq e^\epsilon \times Pr[\mathcal{K}(D_2) \in S].$$

$\epsilon$ is called the privacy budget. One way to guarantee $\epsilon$-differential privacy is by adding Laplacian noise to each component of query results. This noise depends on the sensitivity of the mechanism $\mathcal{K}$, i.e. the largest impact the addition or removal of one single element could have on the result.

In the general case, when applying a mechanism satisfying $\epsilon$-differential privacy many times, we obtain weaker privacy guarantees. Distinct data are often correlated inside the database. For example applying the same mechanism twice on the same integer results in two different noised values and can therefore imply a data leakage. This is known as *sequential composition*.

*Theorem 1:* The composition of mechanisms $\mathcal{K}_i$ satisfying $\epsilon_i$-differential privacy over a database $D$ results in a new mechanism $\mathcal{K}$ satisfying $\left(\sum_i \epsilon_i\right)$-differential privacy.

However in the rare case where we have disjoints databases, there is no more threat of information leakage when applying mechanisms on them. The final privacy guarantee only depends on the worst guarantee offered by these mechanisms. This is known as *parallel composition*.

*Theorem 2:* The composition of mechanisms $\mathcal{K}_i$ satisfying $\epsilon_i$-differential privacy over a set of disjoints databases $D_i$ results in a new mechanism $\mathcal{K}$ satisfying $(\max_i \epsilon_i)$-differential privacy.

## B. Differential privacy for location privacy

Differential privacy can be directly applied to protect location privacy. Authors of [14] have used it to protect trajectories of ships. However, due to the particular nature of geographic locations, there are many ways to add noise to a trajectory. The goal here is to protect the presence or not of a point inside a trace. They introduce three methods to protect these trajectories: adding global noise to the whole trace (by generating a global noise vector), adding noise to each point independently (by generating 2-dimensional noise vectors) or adding noise to each coordinate independently (by applying Laplacian noise to each coordinate). They show that the latter provides the strongest protection but at the cost of a very degraded practical utility.

Another approach to apply differential privacy to location privacy is differentially private data mining inside the traces as proposed in [15]. Authors introduce a way to build a quad-tree and use it to perform a DBSCAN clustering whose result is differentially private. Like in classical differential privacy, the goal here is to protect the result of an aggregate function and not to access directly to the whole data set.

## C. Geo-Indistinguishability

Andrés et al. introduced Geo-Indistinguishability in [6] (Geo-I for short), which is a generalisation of the concept of differential privacy for the protection of location data, represented as a set of secrets $\mathcal{X}$. Instead of revealing its real location, a user report to be in a location inside a set $\mathcal{Z}$ with $\mathcal{P}(\mathcal{Z})$ being the set of probabilities over reported locations $\mathcal{Z}$.

*Definition 2:* A randomized mechanism $\mathcal{K} : \mathcal{X} \longrightarrow \mathcal{P}(\mathcal{Z})$ gives $\epsilon$-geo-indistinguishability w.r.t euclidian distance metric $d_2$ if for every $x, x' \in \mathcal{X}$ and $Z \subseteq \mathcal{Z}$, we have:

$$\mathcal{K}(x)(Z) \leq e^{\epsilon d_2(x,x')} \mathcal{K}(x')(Z).$$

Note that $\epsilon$ coming from Geo-I cannot be compared with $\epsilon$ from differential privacy, whereas they convey the same idea of representing a privacy budget. Authors propose a mechanism to provide Geo-I by adding noise drawn from a 2-dimensional Laplace distribution.

They finally introduce two use cases demonstrating the applicability of Geo-I. The first use case shows how to retrieve features around a user from a LBS. The algorithm they propose guarantees accurate results while preserving the privacy of the user. The idea is to send an obfuscated location satisfying Geo-I to the LBS while extending the retrieval area in order to retrieve all features standing in the original search area around the real user's location with a high probability. As such, the LBS returns a set of candidate results, without learning the effective location of the user. Finally, the mobile client has to filter these candidates to only include those that belong to the original search area, as intended by the user.

The second use case aims at sanitising a static data set containing sensitive location data and studying the impact the sanitisation on the results of statistical queries. In this use case, locations contained in the data set are supposed to be uncorrelated, i.e. they can be sanitised independently without weakening the $\epsilon$-Geo-I privacy guarantee.

As in classic differential privacy, it is known that protecting multiple correlated values with independent noise decreases dramatically the privacy guarantees. We are aware of this fact, yet, as Geo-I is intended to protecting mobile users from curious LBSs (as expressed by the first use case) our objective is to study the effectiveness of the protection provided by Geo-I in practice.

## V. EVALUATION METRICS

We present in this section a set of evaluation metrics that we will use in our study to evaluate the practicability of Geo-I. Specifically, as described in Section II, we aim at evaluating the effectiveness of Geo-I in protecting users POIs. To do so, we first introduce in this section the metrics used to compare the user's real POIs with the obfuscated POIs inferred by the LBS in terms of recall, geographic precision and semantic precision in part V-A. In this part, we further describe the metric used to evaluate the re-identification attack performed by the LBS. Finally, we introduce the metric to evaluate the precision of the results sent by the LBS when receiving requests obfuscated by Geo-I in part V-B.

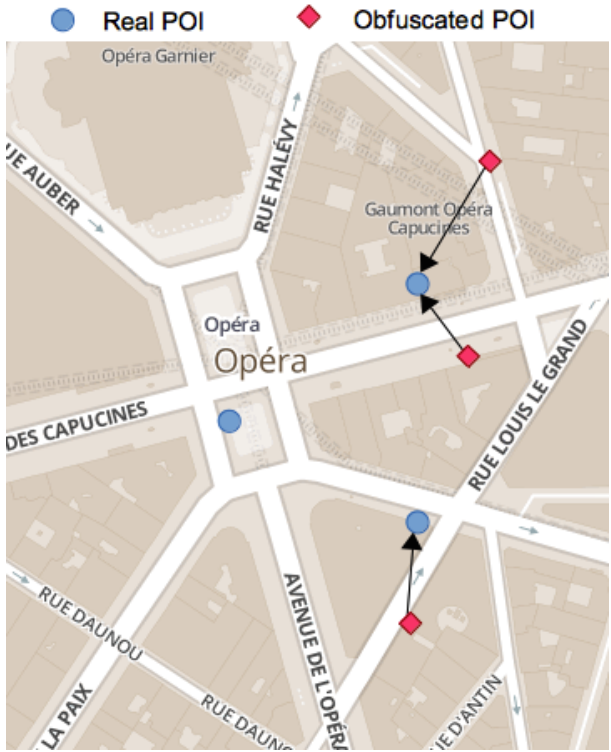

Figure 2. Remapping obfuscated POIs to real ones.

### A. Measuring privacy

To measure the degree of privacy that is practically provided by Geo-I to users, we consider the point of view of a curious LBS, which tries to infer knowledge about them from obfuscated locations they send along with their queries. We assume that the LBS collects obfuscated locations and runs the algorithm described in Section III to infer a set of (obfuscated) POIs associated with the user. The metrics described in this section allows the assessment of the accuracy of obfuscated POIs inferred by the LBS compared to the user's real POIs. Towards this purpose, we use the notations introduced in Table I.

Table I. NOTATIONS REGARDING PRIVACY EVALUATION.

| | |
|---|---|
| $\mathcal{U}$ | Set of all users |
| $\mathcal{C}$ | Set of all spatial locations |
| $\mathcal{L}_u \subset \mathcal{C}$ | Set of real POIs of user $u$ |
| $\mathcal{L}_u^* \subset \mathcal{C}$ | Set of obfuscated POIs of user $u$ |
| $\mathcal{F} \subset \mathcal{C}$ | Set of all existing features |
| $\mathcal{L}^* = (\mathcal{L}_u^* \ \forall u \in \mathcal{U})$ | List of all sets of obfuscated POIs |
| $dist : \mathcal{C}^2 \longrightarrow \mathbb{R}^+$ | Distance between two locations |

The first operation we perform to assess the accuracy of the obfuscated POIs is to remap each of them to the nearest real POI of the same user.

*Definition 3:* The remapping function is defined as follow:
$$remap : \begin{array}{rcl} \mathcal{L}_u^* & \longrightarrow & \mathcal{L}_u, \\ l & \longmapsto & l' \text{ s.t. } dist(l, l') = \min_{p \in \mathcal{L}_u} dist(l, p). \end{array}$$

Figure 2 shows an example of the application of the $remap$ function in the example of Figure 1 in which extracted POIs have been reduced to their centroids. We show the remapping operation with arrows between obfuscated POIs and real POIs. We can already see that a real POI can have zero, one or many obfuscated POIs remapped to it.

**Number of real POIs inferred by the LBS**
Our first metric, i.e., the recall, is inspired from classical metrics in information retrieval. It gives the number of real POIs guessed by the LBS from the obfuscated locations sent by the user. The recall is computed as the ratio of the number of real POIs to which obfuscated POIs have been remapped to and the number of real POIs.

*Definition 4:* The recall is defined for each user as follows:
$$recall : \begin{array}{rcl} \mathcal{U} & \longrightarrow & [0, 1], \\ u & \longmapsto & \frac{|\{remap(l) \ \forall l \in \mathcal{L}_u^*\}|}{|\mathcal{L}_u|}. \end{array}$$

In the example of Figure 2, the recall is equal to $2/3$ because one of the real POIs has no obfuscated POI remapped to it.

The recall gives us a hint about the proportion of POIs that have been retrieved by the LBS and that can be remapped to real POIs, but with no indication about the accuracy of these POIs. Indeed, an obfuscated POI is always remapped to a real POI, but in practice the former could be far from the latter and represent a totally different information about the user. To measure the quality of knowledge inferred by the LBS we define the following two metrics.

**Geographic distance between POIs**
This metric represents the physical distance as given by the $dist$ function between an obfuscated POI and the real one to which it has been remapped.

*Definition 5:* The geographic distance is defined for each obfuscated POI as follows:
$$geographic : \begin{array}{rcl} \mathcal{L}_u^* & \longrightarrow & \mathbb{R}^+, \\ l & \longmapsto & dist(l, remap(l)). \end{array}$$

In the example of Figure 2, the geographic distance corresponds to the length of black arrows.

**Semantic distance between POIs**
This metrics assesses the similarity between the neighbourhood of a POI before and after remapping. Neighbourhood is defined as the top-15 nearest features (e.g., restaurants, doctors, shops) around a given point. To obtain these features we can use the nearest-neighbours search provided by most of LBSs.

*Definition 6:* The function retrieving the $k$ nearest neighbours of a location is defined as follows:
$$top_k : \begin{array}{rcl} \mathcal{C} & \longrightarrow & \mathcal{F}^k, \\ c & \longmapsto & F \text{ s.t. } |F| = k \wedge \nexists f' \notin F, \\ & & \text{s.t. } dist(f', c) < \max_{f \in F} dist(c, f). \end{array}$$

Using this function, we compute the semantic similarity between a real POI and obfuscated one, as the size of the intersection between the $top_k$ features present in the neighbourhood of each of the two POIs out of $k$. To be consistent with the previous metric (which is geographic distance), we prefer to define a semantic distance. As the semantic similarity has a value comprised between 0 and 1, we subtract this similarity to 1 to obtain a distance.

*Definition 7:* The semantic distance is defined for each obfuscated POI as follows:
$$semantic: \mathcal{L}_u^* \longrightarrow [0,1],$$
$$l \longmapsto 1 - \frac{|top_{15}(l) \cap top_{15}(remap(l))|}{|top_{15}(remap(l))|}.$$

**Number of anonymous users re-identified by the LBS**
Using this metric, we aim at evaluating the number of users the LBS is able to re-identify. We consider that a user has been re-identified by the LBS if the latter can unambiguously re-associate its real POIs with its obfuscated ones (or the opposite). To illustrate this attack, we consider the following two scenarios: (1) let us consider an LBS that has collected non-obfuscated location information about a set of users. As such, the LBS can compute their set of real POIs. If some users start using Geo-I obfuscated locations anonymously, we want to measure whether the LBS is able to re-identify them by re-associating their obfuscated POIs with their real ones; (2) let us consider that the LBS has collected obfuscated locations from a set of users. If unluckily the LBS gets anonymous non-obfuscated locations about some users, we want to measure whether the LBS is able to re-identify them by re-associating their real POIs with their obfuscated ones.

To do so we compare two matrices: a matrix of real POIs $R$ where each line is a list of POIs associated with a given user whose identity is known and a matrix of obfuscated POIs $O$ where each line is a list of POIs belonging to the same but unknown user. Then, we compute a distance measurement between each line in $R$ and each line in $O$ and associate each line in the first matrix with the one in the second that has the smallest distance with it.

*Definition 8:* The distance between a set of obfuscated POIs and real POIs of a given user is defined as follows:
$$udist: \mathcal{L}^* \times \mathcal{U} \longrightarrow \mathbb{R}^+,$$
$$L^*, u \longmapsto median(\{\min_{l \in \mathcal{L}_u} dist(l, l') \; \forall l' \in L^*\}$$
$$\cup \{\min_{l' \in L^*} dist(l, l') \; \forall l \in \mathcal{L}_u\}).$$

*Definition 9:* The re-identification function which associates to a list of obfuscated POIs the most probable user it comes from is defined as follows:
$$uassoc: \mathcal{L}^* \longrightarrow \mathcal{U},$$
$$L^* \longmapsto u \text{ s.t. } udist(L^*, u) = \min_{u' \in \mathcal{U}} udist(L^*, u').$$

The amount of users re-identified by the LBS is thus the percentage of lines in $R$ that have been assigned with the right line in $O$.

*Definition 10:* The re-identification rate is defined as follows:
$$reident = \frac{\sum_{u \in \mathcal{U}} \begin{cases} 1 \text{ if } uassoc(\mathcal{L}_u^*) = u, \\ 0 \text{ otherwise} \end{cases}}{|\mathcal{U}|}.$$

### B. Measuring precision of LBS results

Maintaining a high accuracy has a cost. We use the mechanism presented in [6] where results are retrieved within a larger area than originally intended by the user. Then, the application running on the user's device filters back results coming from the LBS and only presents to the user those that really matches its original query. The modified query is such that accurate results should be retrieved with a given probability, specified by the user. Obviously, increasing this probability or the level of privacy guaranteed by Geo-I increases the amount of useless results the user has to filter. We thus measure the precision of results returned by the LBS when receiving an obfuscated location from the user by comparing the number of useless results with the total number of retrieved results. Towards this purpose, we use the notations introduced in Table II.

Table II. NOTATIONS REGARDING PRIVACY EVALUATION.

| | |
|---|---|
| $\mathcal{C}$ | Set of all spatial locations |
| $\mathcal{F} \subset \mathcal{C}$ | Set of all existing features |
| $res_q : \mathcal{C} \longrightarrow \mathcal{F}^n$ | Results for a query $q$ around a given location |
| $obf_\epsilon : \mathcal{C} \longrightarrow \mathcal{C}$ | Function obfuscating a location w.r.t. $\epsilon$-Geo-I |

*Definition 11:* The precision for a given privacy level $\epsilon$ and query $q$ is defined for each location as follows:
$$precision_{\epsilon,q}: \mathcal{C} \longrightarrow [0,1],$$
$$c \longmapsto 1 - \frac{|res_q(obf_\epsilon(c)) - res_q(c)|}{|res_q(obf_\epsilon(c))|}.$$

The precision we compute has a direct consequence on the cost incurred to mobile devices as each useless retrieved result impacts the performance of the latter (e.g., in terms of battery). In this paper, we do not measure the impact of using Geo-I in terms of battery consumption but we plan to carry out this study in our future works.

## VI. EXPERIMENTAL RESULTS

### A. Data sets

We use two real-life data sets to conduct our experiments. The first one is the well-known mobility traces from the San Francisco cabs [8]. These traces have been collected over a month by 536 taxis in the San Francisco Bay Area. Each taxi was equipped with a GPS tracker and locations were all collected to a central server. The resulting data set is composed of 11 219 955 locations, with on average 21 000 locations per taxi.

The second one is the Geolife data set [9], [16], [17]. It was collected by Microsoft Research Asia over four years by 182 users. It is not restricted to people working hours but instead follows people during the whole day. This data set includes 25 050 848 locations, but the variance of the number of locations per user is very high: some people have been tracked during the whole period of the experiment whereas others have only contributed for a few days. We have filtered the data set to keep only days with more than 480 locations and users with at least 30 days. After processing, the final data set contains 5 476 442 locations for 61 users.

### B. Initial extraction of POIs

The POIs extraction algorithm presented in Section III requires to define the $minTime$ and $maxDistance$. We choose $minTime = 1$ hour and $maxDistance = 250$ meters in order to capture typical activities occurring in areas of the size of a small neighbourhood in an urban environment. A place was considered as a POI if at least two points in the data set satisfies $minTime = 1$ and $maxDistance = 250$ during the length of the data set. The same parameters have been used with the two data sets. Sum-up of this section is in Table III.

Table III.  SUMMARY OF POIs EXTRACTION SETTINGS.

| | |
|---|---|
| $minTime$ | 1 hour |
| $maxDistance$ | 250 meters |
| $minPts$ | 2 |
| #POIs on SF cabs | 1111 |
| #POIs on Geolife | 258 |

With the San Francisco data set, we identify 1111 POIs which gives an average of two POIs per driver. For the Geolife data set, we obtain 258 POIs which corresponds to an average of four POIs per user. We can check the correctness of the POIs by attaching semantic information to them using reverse geocoding, like it has been done by authors of [1], [2] to validate their results. With the San Francisco data set we can identify some well-known hotspots such as the taxi company's parking or the San Francisco international airport. With the Geolife data set, we can again find out semantic information associated with those POIs like university and residential areas.

### C. Protection mechanism parametrization

Geo-I described in Section IV is defined for a privacy level $\epsilon$. We have ran our experiments for three values of $\epsilon$. We used values similar to the ones used in the paper introducing Geo-I [6] where $\epsilon = \ell/r$, $\ell$ being the level of privacy we want within a radius $r$ (and then decreasing proportionally outside this radius). The three different levels of privacy are given in Table IV.

Table IV.  VALUES OF $\epsilon$.

| $\epsilon = \ell/r$ | $\ell$ | $r$ (meters) | Characterisation |
|---|---|---|---|
| 0.00139 | $\ln(2)$ | 500 | Strong privacy |
| 0.00358 | $\ln(6)$ | 500 | Medium privacy |
| 0.00693 | $\ln(4)$ | 200 | Weak privacy |

### D. Extracting POIs from obfuscated data set

Geo-I uses a randomized mechanism: the results obtained are not deterministic. Therefore, each test has been ran 10 times against 10 independently obfuscated data sets. Throughout this section and Section VI-E, the results we present are the average values obtained over these 10 runs. 10 different runs seems a reasonable and practicable value to obtain accurate results while keeping a bounded execution time.

The attacker knows the protection mechanism and its corresponding privacy parameter $\epsilon$. He is therefore able to set up his attack with optimal parameters to retrieve the "best" POIs w.r.t. our metrics. The location is the only data affected by the perturbation, there is no reason to modify the time threshold $minTime$ or the minimum number of points $minPts$ in our clustering algorithm. We only modify the distance threshold $maxDistance$ when working with the obfuscated data set.

Figure 3 shows how the recall varies with the distance threshold. Beware of the scale of the horizontal axis which is not the same on each graph, because we have zoomed on the most interesting part. As expected we observe the recall is increasing with the distance threshold. At a fixed threshold, recall is better when the guaranteed privacy is weaker. Geo-I tends to create locations outside a given cluster. By increasing the distance threshold we limit this effect and hence include again obfuscated locations that were lost ouside a cluster. The drawback is than if we choose a threshold too high, results will be imprecise resulting in high geographic and semantic distances.

We can now choose an optimal threshold which is a threshold for which we want a high recall and low geographic and semantic distances. We choose the minimum threshold for which the recall is higher than 70 % when it was possible. Because the experiment was ran for discrete values of thresholds sampled every 100 meters, we give them with a 100 meters precision. We present in Table V the distance thresholds for which we have a recall of more than 70 %. Note that this goal was not reachable when working with Geolife at the highest privacy level.

Table V.  "OPTIMAL" DISTANCE THRESHOLDS.

| | SF cabs | Geolife |
|---|---|---|
| $\epsilon = 0.00139$ | 2000 meters | 2500 meters |
| $\epsilon = 0.00358$ | 1000 meters | 1200 meters |
| $\epsilon = 0.00693$ | 700 meters | 600 meters |

### E. Privacy guarantees

In all this section, all extractions of POIs on the obfuscated data set use final distance thresholds presented in Table V.

**Recall rate**
We chose our parameters to guarantee a certain recall as shown in previous section. Hence there will be no surprise with recall rates across all users we present in Table VI.

Table VI.  RECALLS FOR DIFFERENT VALUES OF $\epsilon$.

| | SF cabs | Geolife |
|---|---|---|
| $\epsilon = 0.00139$ | 71.01 % | 60.57 % |
| $\epsilon = 0.00358$ | 71.54 % | 70.56 % |
| $\epsilon = 0.00693$ | 73.31 % | 71.94 % |

**Geographic distance**
Cumulative distribution of geographic distance across all users for different values of $\epsilon$ is shown in Figure 4. The lowest level of privacy offers weak privacy guarantees: 80 % of the POIs are within a 200 meters range of a real POI with both data sets. For the strongest level of privacy, we have only 25 % (San Francisco data set) and 20 % (Geolife data set) of the POIs within the 200 meters range. Please note that due to the exponential distribution used to draw noised locations, there is a long tail of obfuscated POIs having a high geographic distance that is not shown in the graph to keep it readable.

**Semantic distance**
Cumulative distribution of semantic distance across all users is shown in Figure 5. At the lowest level of privacy, 70 % (San Francisco data set) and 80 % (Geolife data set) of POIs have a semantic distance of less than 10 %, whereas with the strongest privacy level only 15 % (San Francisco data set) and 45 % (Geolife data set) are in that case.

**Re-identification of users**
We finally have conducted the re-identification attack and present results in Table VII. Re-identification rates are bad with the SF cabs data set while they are good with the Geolife data set. It is due to the fact that the Geolife data set contains data from people that have regular mobility patterns (think to your daily home-work-home journey) and hence more subject to re-identification than taxi drivers that all share the same POIs because of their professional duties (e.g., they all often go the San Francisco airport, park their car in the same company parking, etc.). Accordingly with a set of studies about uniqueness of human mobility like [18] we are hence able to re-identify a large proportion of individuals.

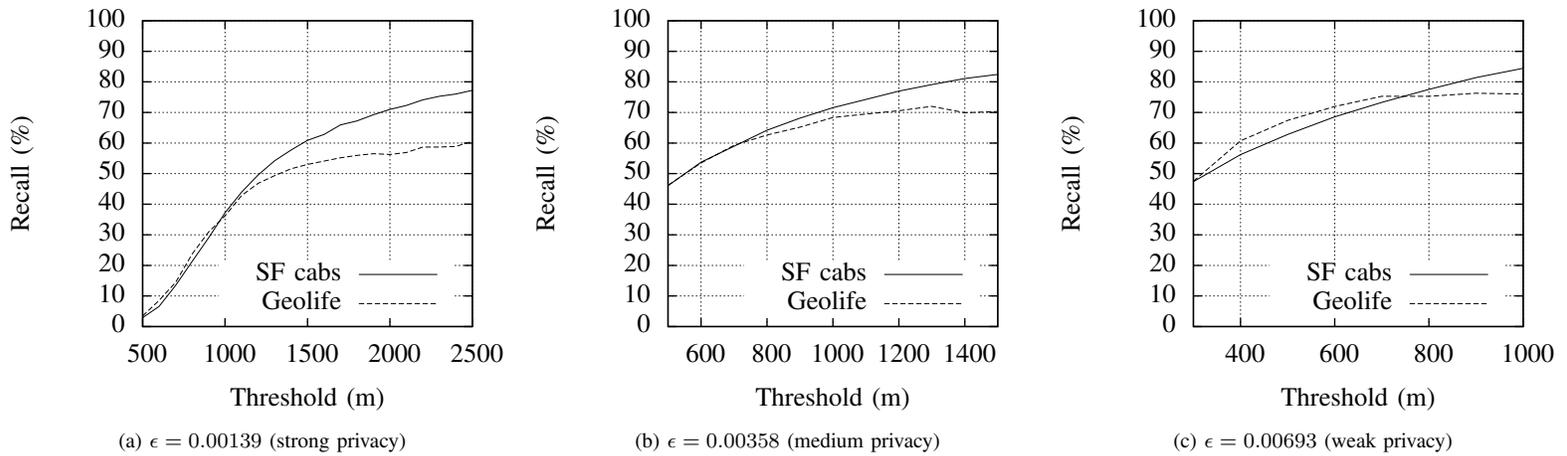

Figure 3. Evolution of recall rate for different values of $\epsilon$.

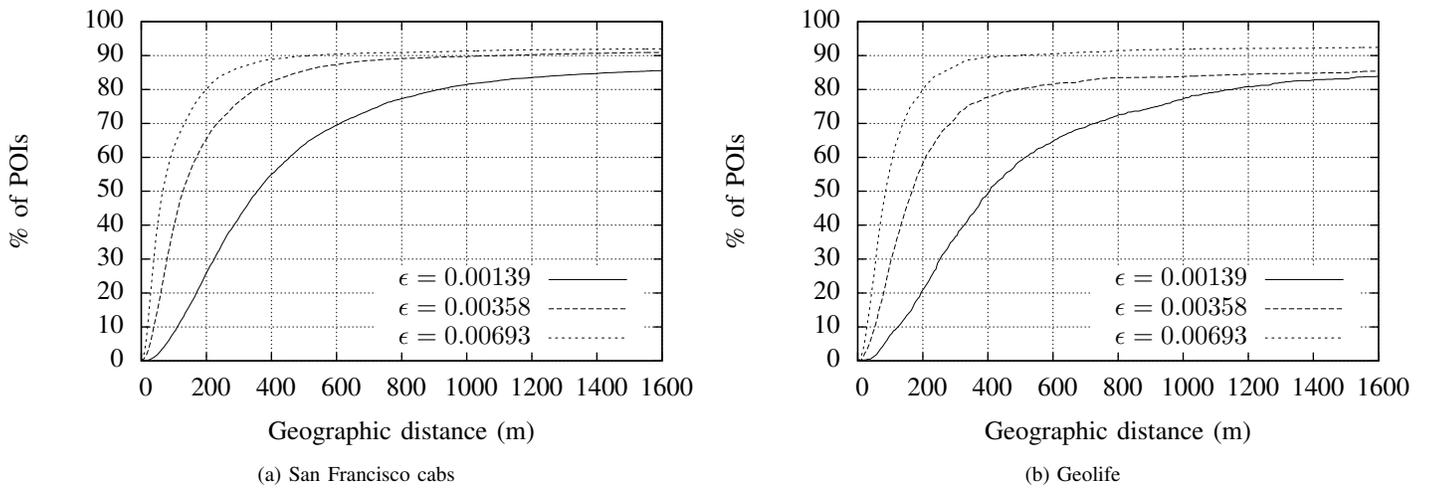

Figure 4. Cumulative distribution of geographic distance.

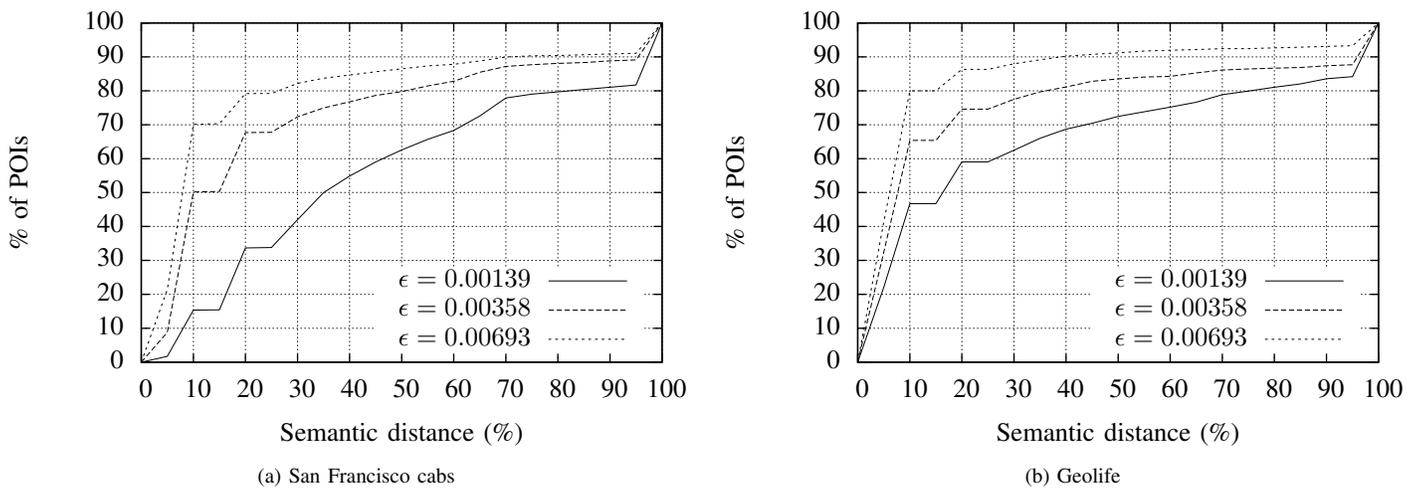

Figure 5. Cumulative distribution of semantic distance

Table VII. RE-IDENTIFICATION RATES FOR DIFFERENT VALUES OF $\epsilon$.

|  | SF cabs | Geolife |
|---|---|---|
| $\epsilon = 0.00139$ | 5.79 % | 63.04 % |
| $\epsilon = 0.00358$ | 8.12 % | 82.90 % |
| $\epsilon = 0.00693$ | 9.66 % | 89.63 % |

*F. Precision with the obfuscated data set*

Instead of querying an LBS online each time it was needed, we used a local database containing a set of features coming from the OpenStreetMap project[3] around the San Francisco Bay Area. We can query aggressively our LBS without worrying about fees or network latency. Our experimental setting uses a typical query: "*find all restaurants 500 meters around me*". Measures have been done over 100 points sampled from the San Francisco cabs mobility trace and while guaranteeing an accuracy of 85 %, i.e., at least 85 % of the real results should be retrieved, among with useless results. Results presented in Figure 6 are average values. Figure 6 describes the evolution of the precision when $\epsilon$ varies. Precision logically increases with $\epsilon$ but always stays relatively low. For the strongest level of privacy, we have only a precision of 8 %, which means 92 % of retrieved results are useless. Even at the weakest level of privacy we only have a precision of 43 %.

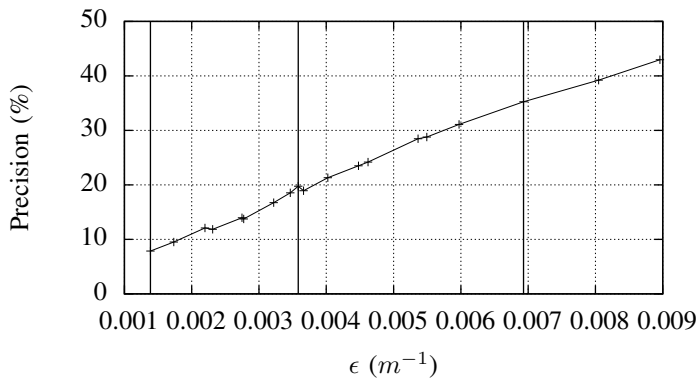

Figure 6. Precision as a function of $\epsilon$.

## VII. RELATED WORK

Shokri et al. in [19] developed a framework to quantify location privacy based, like us, on the success of inference attacks. They have implemented their framework as a C++ tool and tested it against the San Francisco cabs data set.

[3]http://www.openstreetmap.org

However they do not propose a way to evaluate the precision or performance of a protection mechanism when used in a practical use case.

Micinski et al. in [20] conducted a practical experimentation on the effects of truncating the location (i.e., remapping it to the nearest location on a fixed-size grid) on the utility of results when using some Android applications. They have shown that the utility remains stable for a truncation under 5 km. However, maximum truncation that can be applied without losing too much utility depends on the population density of the surrounding region.

Gambs et al. in [21] and [2] have made a large number of attacks against location privacy by using mobility traces. They made it clear that it was possible to deanonymise the San Francisco cabs mobility trace and ultimately deanonymize the whole trace. They used similar clustering techniques to ours, although they do not perform an extensive study of protection mechanism as we did in this paper.

Krumm studied threats and protection mechanisms for location privacy in [1]. He tried to find homes from mobility traces using different heuristics. The median error was of 60.7 meters with its best algorithm and it ultimately succeed in associating the correct name to 5 % of mobility traces. The author then analysed the impact of protection mechanisms like spatial cloaking, Gaussian noise and inaccuracy (where data is mapped on a grid with fixed precision). In particular, he was able to determine how much noise is necessary to hide successfully most of homes from a data set.

A set of works tries to characterise the uniqueness of human mobility. Authors of [18] show than only four points are sufficient to uniquely identify a trace among other traces. Authors of [22] extensively studied the uniqueness of the home/work couple with different granularity levels. At the scale of a census block, there is basically no privacy because this pair is likely to be unique. At the scale of a census tract, it becomes unique for 5 % of people but still offers little privacy for the others.

## VIII. CONCLUSION

We have evaluated a differentially-private mechanism for location privacy. Results show that it is possible to improve privacy by using a location-privacy protection mechanism. However it is still possible to infer a large part of a user's POIs with a reasonable precision. In our experiments only the strong and medium levels of privacy really achieve the goal of hiding one's POIs, the weakest level leading to a very precise re-identification of POIs. But we also show it has a non negligible cost: precision of retrieved results is very low with the highest privacy level and is likely to lead to degraded performance on a mobile device with limited computational power. It seems a trade-off between privacy and precision is hard to obtain with state-of-the-art techniques.

While analysing results, it appears to us that privacy would be better guaranteed if a user only moved in a not too large area. If all his frequent mobility patterns are done within a small area, obfuscated POIs from an obfuscated trace become hard to distinguish because added noise will tend to merge all these POIs into a unique POI. It still allows us to determine a neighbourhood where this user evolves, but we will not be able to identify his home and his work place and then re-identify him.

Differentially private mechanisms usually noise only one information: the location. Two other pieces of information are attached to each element of a mobility trace: a timestamp and a link with a unique user. This gives us an advantage and we use it when extracting POIs. We want in particular to study the impact of the time when extracting POIs and how it is possible to hide it. In the same time we are interested in the feasibility of unlinkability between user queries and how this affects threats studied in this paper.

The efficiency of our attack is related to the fact the attacker knows the exact privacy level that is used to protect a mobility trace. Knowing that, he is able to adapt the parameters of his attack to have better results. We are interested in studying in which measure it is possible for an attacker to find out the level of privacy $\epsilon$ that has been applied on a trace by analysing the entropy of a trace. Thereafter a counter-measure can be for a user to vary the amount of noise dynamically depending on the environment (urban/countryside place, users' density).

ACKNOWLEDGMENT

This work was supported by the LABEX IMU (ANR-10-LABX-0088) of Université de Lyon, within the program "Investissements d'Avenir" (ANR-11-IDEX-0007) operated by the French National Research Agency (ANR).